\documentclass[apj]{emulateapj}

\usepackage{graphicx}
\usepackage{calrsfs}
\usepackage{amsbsy}
\usepackage{amssymb}
\usepackage{amsmath}
\usepackage[hypertex]{hyperref}

\slugcomment{ApJ, to be submitted}

\shorttitle{Beaming Pattern of External Compton Radiation}
\shortauthors{S. R. Kelner et al.}

\begin{document}
\def\b{\boldsymbol}
\def\p{\partial}
\def\vt{\vartheta}
\def\e{\epsilon}
\def\ts{\textstyle}
\def\k{\varkappa}
\def\d{\delta}
\def\dd{\mathcal{D}}
\def\t{\tilde}
\def\be{\begin{equation}}
\def\ee{\end{equation}}
\def\ba{\begin{eqnarray}}
\def\ea{\end{eqnarray}}

\title{The beaming pattern of External Compton Emission from relativistic outflows}

\author{S. R. Kelner\altaffilmark{1,2}}
\affil{$^1$ Max-Planck-Institut f\"ur Kernphysik, P.O. Box 103980, 69029
Heidelberg, Germany}
\affil{$^2$ National Research Nuclear University (MEPHI), Kashirskoe shosse
31, 115409 Moscow, Russia}
\email{Stanislav.Kelner@mpi-hd.mpg.de}
\author{E. Lefa\altaffilmark{1}}
\affil{$^1$ Max-Planck-Institut f\"ur Kernphysik, P.O. Box 103980, 69029
Heidelberg, Germany}
\author{F.M. Rieger\altaffilmark{1}}
\affil{$^1$ Max-Planck-Institut f\"ur Kernphysik, P.O. Box 103980, 69029
Heidelberg, Germany}
\author{F. A. Aharonian\altaffilmark{3,1}}
\affil{$^1$ Max-Planck-Institut f\"ur Kernphysik, P.O. Box 103980, 69029
Heidelberg, Germany}
\affil{$^3$ Dublin Institute for Advanced Studies, 31 Fitzwilliam Place, Dublin
2, Ireland}

\begin{abstract}
The beaming pattern of radiation emitted by a relativistically moving source like jets in microquasars, 
AGN and GRBs, is a key issue for understanding of acceleration and radiation processes 
in these objects. In this paper we introduce 
a formalism based on a solution of the photon transfer equation to study the 
beaming patterns for emission produced by electrons accelerated in the jet and
upscattering photons of 
low-energy radiation fields of external origin (the so-called External Compton
scenario). 
The formalism allows us to treat 
non-stationary, non-homogeneous and anisotropic distributions of electrons, but
assuming 
homogeneous/isotropic and non-variable target photon fields. 
We demonstrate the non-negligible impact of the anisotropy in the electron
distribution on angular and spectral characteristics of the EC radiation.

\end{abstract}


\maketitle

\section{Introduction}
The inverse Compton scattering (ICS) of relativistic electrons is one of the
major radiation mechanisms in high energy astrophysics. Since the Compton
cooling time of electrons decreases linearly with energy, this channel becomes
especially prolific in the gamma-ray band.
The universal presence of dense radiation fields makes ICS an effective 
gamma-ray production mechanism in various astronomical environments, in
particular in sources containing relativistic outflows - microquasars, AGN,
GRBs, {\it etc.} The dramatically enhanced fluxes, the shift of the spectral
energy distribution towards higher energies, and the reduction of characteristic
timescales are distinct features of the Doppler boosted radiation produced in a
relativistically moving source with a Doppler factor $\dd \gg 1$. In
relativistic outflows, ICS can be realized, depending on the origin of target
photon fields, in two different scenarios. 

In the so-called Synchrotron-Self-Compton (SSC) model the synchrotron photons
of relativistic electrons constitute the main target for Compton scattering of
the same electrons. In its simplified homogeneous one-zone version, the SSC
model assumes that a spherical source (a ``blob'') 
filled with an isotropic electron population and a random magnetic field, moves 
towards the observer with a constant Doppler factor $\dd$. 
The beaming pattern 
for emission of  a relativistically moving source 
derived by \cite{Lind85}, can 
be directly applied to the Synchrotron and SSC components of radiation. 

In the second scenario, the seed photon field for ICS is dominated 
by external radiation, i.e. by low-energy photons produced outside the 
moving source of relativistic electrons. In this, the so-called External Compton
(EC) model
the beaming pattern of the inverse Compton radiation differs significantly from
the 
beaming pattern of the synchrotron and SSC components \citep{Dermer95}. 

The characteristics of EC radiation of a relativistically moving source can be
calculated using two different approaches: 

(i) transforming first the external radiation photon field to the jet frame,
calculating in the same frame the characteristics of the high energy photon
(the outcome of ICS), 
and finally transforming the latter back to the observer frame; 

(ii) transforming the electron distribution from the jet frame to the observer 
frame, 
and then calculating the spectrum of IC photons directly in the observer 
frame. 

Using the first approach, \cite{Dermer95} has derived the EC beaming pattern in
the Thomson limit assuming a power-law distribution of electrons. 
\cite{Kirk01} using the second approach have extended this result to the
general case of Compton scattering, including the Klein-Nishina regime. 
In both treatments, the distribution of relativistic electrons has been assumed
to be isotropic and homogeneous. In the zeroth approximation, this could be a
reasonable assumption, and thus can be in principle applied to the 
interpretation of gamma-ray observations of many blazers. However, in some 
other cases one cannot exclude significant 
deviations of distributions of electrons from homogeneous and isotropic
realizations. 
The non-stationary treatment of the problem is another issue which has not been 
addressed so far. 

In this paper, we develop a new approach which provides a strict formalism
for the treatment of the beaming pattern for EC radiation. Namely, we solve the
photon transfer equation which allows us to  the beaming pattern in a
concise way, and, more importantly, to extend  the formalism to more general
(non-stationary and anisotropic) case of electron distributions, but assuming
isotropic, homogeneous and isotropic distribution of the seed (target) photon
fields. For demonstration of the potential of the proposed formalism, we
examine how anisotropy in the electron distribution affects the 
the energy spectrum and the angular distribution of  EC
radiation.

\section{The beaming of external Compton emission}\label{BEC}

\subsection{The photon transfer} \label{photon_ transf}

In two close points located on the line of sight, the specific intensity (the
spectral radiance) $I$ and emissivity $j$ are related as (see e.g. 
\citealt{Rybicki})
\be\label{ec1}
I(s)-I(s-ds)=j\,ds\,,
\ee
where $s$ is the distance along the line of sight. It is assumed than one can
neglect 
the scattering and absorpion of radiation during its propagation. It is
convenient 
to consider $I$ and $j$ as functions of the radius-vector $\b r$, the time
$t$, and the wave-vector of the photon $\b k$, i.e. $I\equiv I(\b k,\b r,t)$,
and 
$\b k$, i.e. $s\equiv s(\b k,\b r,t)$. For simplicity, hereafter we will use
the 
system of units in which the speed of light $c=1$ and the Planck constant 
$\hbar=1$. 
Also, instead of $I$ and $j$ we will use the distribution function of photons 
$g=I/\e^3$ and the source function $Q=j/\e^3$, where $\e=|\b k|$ is the photon
energy. 
The function $g(\b k,\b r,t)\,d^3k\,d^3r$ describes the number of photons in
the volume element
$d^3k\,d^3r$ of the phase space at the moment $t$, while $Q(\b k,\b
r,t)\,d^3k\,d^3r\,dt$ 
is the number of photons in the momentum interval $d^3k$, 
emitted during the time interval $(t,t+dt)$ from the volume 
$d^3r$ of the source located at the point $\b r$. With these new notations,
Eq. (\ref{ec1}) can be written in the form
\be\label{ec2}
g(\b k,\b r,t)-g(\b k,\b r-\b n ds,t-ds)=Q(\b k,\b r,t)\,ds\,,
\ee
where $\b n=\b k/|\b k|$ is a unit vector along the photon momentum. 
Using Eq.~(\ref{ec2}), it is easy to derive the following relation
\be\label{ec3}
g(\b{k},\b{r},t)=\int^{\infty}_{0}\! ds\,
Q\big(\b{k},\b r-\b ns,t-s\big)\,,
\ee
which expresses  $g$ through $Q$. 
For calculations of the flux one should integrate $g$ over the directions of
the vector
$\b n$. This determines the total number of photons of a given energy which 
reach 
the point $\b r$ at the moment $t$, 
\be\label{ec4}
\t g(k,\b r,t)=\int\! g(\b k,\b r,t) d\Omega_{\b n}\,.
\ee

For integration over $d\Omega_{\b{n}}$, it is convenient to write 
Eq.~(\ref{ec3}) in the form
\be\label{ec5}
g(\b{k},\b{r},t)=\!\int^{\infty}_{0}\!\! ds\!\int\! d^3r_0^{}\,
Q\big(\b k_r,\b r_0^{},t-s\big)\,\d(\b r-\b r_0^{}-\b n s),
\ee
where $\d(\b r-\b r_0^{}-\b n s)$ is the three-dimensional $\d$-function, and
$\b k_r=\e(\b
r-\b r_0^{})/|\b r-\b r_0^{}|$. 
After a simple integration of Eq.~(\ref{ec5}) over $d^3r_0^{}$, we apparently
would return to Eq.~(\ref{ec3}).
In the integrand of Eq.~(\ref{ec5}), the vector $\b n$ enters only in the
argument of the
$\d$-function, therefore for determination of $\t g$, one should calculate 
the integral 
\be\label{ec6}
X \equiv \int\!\d(\b r-\b r_0^{}-\b n s)\,d\Omega_{\b n}\,.
\ee
Using the equation
\be
\d(\b r-\b r_1^{})=\frac{1}{r^{2}\,\sin\theta}\d(r-r_1^{})
\,\d(\theta-\theta_1^{})\,\d(\phi-\phi_1^{})\,,
\ee
where $r,\,\theta,\,\phi$ ($r_1,\,\theta_1,\,\phi_1$) are the spherical
coordinates of the point 
$\b r$ ($\b r_1$), we find 
\be\label{ec6}
X=\frac{\d(|\b{r}-\b{r}_0|-|s|)}{|\b{r}-\b{r}_0|^2}\,.
\ee

Then the function $\t g$ becomes
\[
\t g(k,\b{r},t) =\int_{0}^{\infty}ds\!\int\! d^3r_0\,
\frac{Q(\b k_r,\b r _0,t-s)}{|\b r-\b r_0|^2}
\]
\be\label{ec7}
\times\d(|\b{r}-\b{r}_0|-s)\,,
\ee
and the integration over $s$ results in
\be\label{ec8}
 \t g(k,\b{r},t)=\int\! d^3r_0\, \frac{Q(\b k_r,\b r_0,t-|\b r-\b r_0|)}
{|\b r-\b r _0|^2}\,.
\ee
This equation has a simple physical meaning and can be also obtained from 
heuristic considerations. From the definition of the function 
$Q$, it follows that 
\be\label{ec9}
\frac{Q\,d^3r_0}{|\b r-\b r_0|^2}\,\e^2\,d\e
\ee
is the number of photons in the unit volume at the point 
$\b r$ and in the energy interval $d\e$, emitted by a point-like source located
at the point $\b r_0$. Note that due to the delay, the number of photons at the
instant $t$ is determined by the source function at the moment $t-|\b r-\b
r_0|$. In the case of an extended source, for calculation of the number of
photons one should use the sum of Eq.~(\ref{ec9}) type expressions. The result
will be the quantity $\t g\,\e^2\,d\e$. Replacing the sum by the integral, we
would arrive at Eq.~(\ref{ec8}).

Eq.~(\ref{ec8}) is correct for any $\b r$ and $t$. In order to
calculate the photon distribution at a distance that significantly exceeds the
characteristic size of the source $R_0$, let's define the origin of coordinates
somewhere inside the source. Then for  $r \gg R_0$, one can set $\b n_r=\b r/r$
and $|\b r-\b r_o|=r-(\b n_r\b r_0)$. Neglecting in the denominator $(\b n_r\b
r_0)$ compared to $r$, we obtain 
\be\label{gtilde}
\t g(k,\b{r},t)=\frac1{r^2} \int\! d^3 r_0\,
Q\big(\b k,\b r_0,t-r+(\b n_r\b r_0)\big)\,,
\ee
where $\b k=k\b n_r$. This means that all photons detected by a distant
observer travel essentially in the direction of $\b{r}$. Therefore below we
will not distinguish between the vectors $\b n=\b k/k$ and $\b n_r=\b r/r$,
i.e. $\b n_r=\b n$. On the other hand, we should note that in general one
cannot neglect the last term in the argument $t-r+(\b n\b r_0)$. The
possibility of such an approximation is determined not by the relation of $r$
to $(\b n\b r_0)$, but by the speed of variation 
of the function $Q$ during the time $(\b n\b r_0)$.

\subsection{The beaming pattern} \label{beam_patt}

Let us consider the collision of ultrarelativistic electrons and soft photons. It
is convenient to describe the ensemble of electrons by the distribution function
in the phase space  $f(\b p,\b r,t)$. By definition, the quantity $f_e(\b p,\b
r,t)\, d^3p\,d^3r$ is the number of electrons located at the moment of time $t$
in the phase volume element $d^3p\,d^3r$. Let's define $n_{\rm ph}(\e_{\rm
ph})\,d\e_{\rm ph}$ as the number of photons of energy  $\epsilon_{\rm ph}$ 
within the interval
$d\e_{\rm ph}$ in the unit volume. We assume that the distribution of these
target photons is homogeneous, isotropic and stationary. Then the high energy
photons appearing due to ICS are described by the source function
\be\label{ex_com1}
Q(\b k,\b r,t)=\int\! w(\b p,\e_{\rm ph},\b k)\,f_e(\b p,\b r,t)\,
n_{\rm ph}(\e_{\rm ph})\,d^3p\,d\e_{\rm ph} ,
\ee
where $w(\b p,\e_{\rm ph},\b k)$ is the probability of scattering averaged
over the directions of the soft photons. According to Eq.~(\ref{gtilde}), 
the corresponding expression for $\t g$ is given by 
\[
\t g(k,\b r,t)=\frac1{r^2}\int\! d^3r_0\,d^3p\,d\e_{\rm ph}\,
w(\b p,\e_{\rm ph},\b k)
\]
\be\label{com2}
\times f_e(\b p,\b r_0,t-r+\b n\b r_0)\,
n_{\rm ph}(\epsilon_{\rm ph})\,.
\ee

All quantities in Eq.~(\ref{com2}) are relevant to a reference system $K$ (the
observer system). Let's assume that the accelerated electrons belong to a blob
which moves with a relativistic speed $V\sim 1$ and Lorentz factor
$\Gamma=1/\sqrt{1-V^2}$. In order to 
determine the impact of the blob's bulk motion on the energy distribution of 
secondary photons, we introduce a co-moving coordinate system 
$K'$, and define $f'(\b p',\b r',t')$ as the distribution function of
electrons in the blob, i.e. in the $K'$ system. Below all quantities in the
$K'$ system will be indicated by the prime symbol. Note that the distribution
function in the phase space is a relativistic invariant (\cite{Landau2},
\cite{Rybicki}), i.e. 
\be\label{com3}
f(\b p,\b r,t)=f'(\b p',\b r',t')\,,
\ee
where the ``primed'' and ``unprimed'' variables are connected via Lorentz
transformations.

If we assume (without a loss of generality) that the source is moving
along the $z$-axis, then Eq.~(\ref{com3}) can be written in the form
\be\label{com4}
f(\b p,x,y,z,t)=f'\big(\b p',x,y,\Gamma(z-Vt),\Gamma(t-Vz)\big),
\ee
where
\be\label{com5}
p'_x=p_x^{}\,,\;\;p'_y=p_y^{}\,\,p'_z=\Gamma(p_z^{}-VE_e)\,.
\ee
In Eq.~ (\ref{com2}) the function 
$f$ is given at an instant delayed in time. Replacing in 
Eq.~(\ref{com4}) $t$ by  $t-r+\b n\b r_0$, one finds
\[
 f(\b p,x_0,y_0,z_0,\tau+n_x x_0+n_y y_0+n_z z_0)=
\]
\[
f'\big[\b p',x_0,y_0,\Gamma\big(z_0-V(\tau+n_x x_0+n_y y_0+n_z
z_0)\big),
\]
\begin{equation}\label{com6}
\Gamma\big(\tau+n_x x_0+n_y y_0+n_z z_0-Vz_0\big)\big],
\end{equation}
where $n_x, n_y,n_z$ are the components of the unit vector $\b n$. Note that
for
simplicity we have set the retarded time as $\tau\equiv t-r$.

Let's introduce the new variables of integration in a way that the space
arguments
of the function $f'$ can be written in the form 
$(x'_0,y'_0,z'_0)$. Apparently, for this we should set 
\ba
&x'_0=x_0\,,\quad y'_0=y_0\,,&\nonumber\\
&z'_0=\Gamma\big(z_0^{}-V(\tau+n_{x}x_0+n_{y}y_0+n_{z}z_0)\big)\,.&\label{com7}
\ea
As it follows from Eq.~(\ref{com7}), $d z'_0=\Gamma(1-Vn_{z})dz_0$; therefore
the volume element is transformed according to
\be\label{com8}
d^{3}r_0=\frac{d^3r'_0}{\Gamma(1-V n_z)}=\dd d^3 r'_0\,,
\ee
where
\be\label{com9}
\dd=\frac1{\Gamma(1-V n_z)}=\frac1{\Gamma(1-\b n\b V)}
\ee
is the Doppler factor. Note that the transformation given by Eq.~(\ref{com8}) 
differs from the standard Lorentz transformation. At $\dd>1$, we have an
increase (but not a contraction as in the case of Lorentz transformation) 
of the volume by a factor of $\dd$.

The time-argument in the right part of Eq. (\ref{com6}) with respect to
the new variables $(x'_0,y'_0,z'_0)$ becomes
\be\label{com10}
\dd\cdot\left(\tau+n_x x'_0+n_y y'_0+\Gamma(n_z-V)z'_0\right).
\ee
Let's introduce a unit vector along $\b k'$.
Using Eq.(\ref{com5}), we obtain the following 
well-known expressions for the aberration of light:. 
\be\label{com11}
n'_x=\dd n_x^{},\;\;\;n'_y=\dd n_y^{},\;\;\;n'_z=\dd\Gamma(n_z^{}-V)\,. 
\ee
Then the time-argument in Eq.~(\ref{com6}) can be written as
\be
\dd \tau+(\b n'\b r'_0)\,.
\ee
As a result, we find that the function $\t g$ in the system $K$ 
is expressed via $f'$  in the system $K'$ as 
\[
 \t g(k,\b r,t)=\frac{\dd}{r^2}\int\! d^{3}r'_0\,d^3p\,d\e_{\rm ph}\,
w(\b p,\e_{\rm ph},\b k)
\]
\be\label{com12}
\times f'_{e}(\b p',\b r'_0,\dd\tau+\b n'_{r}\b r'_0)\,
n_{\rm ph}(\epsilon_{\rm ph})\,.
\ee
Remarkably, the function $f'$ is not constrained by any condition; the
distribution of electrons in the comoving frame can be {\it non-stationary}, 
{\it non-homogeneous}, and {\it anisotropic}. We should note that, in the case
of a homogeneous, isotropic and stationary target photon field, the homogeneity
of electrons becomes irrelevant, i.e. Eq.~(\ref{com12}) does not depend on the
spatial distribution of electrons. 

Eq.~(\ref{com12}) can be significantly simplified after the following
approximations. Let's assume that the distribution function of electrons in the
comoving frame $K'$ is stationary and isotropic, i.e. $f'=f'(E'_e,\b r'_0)$,
where $E'_e$ 
is the electron energy in the $K'$ frame. 
The energy $E_e$ in the $K$ frame and $E'_e$ are related as 
$E'_e=E_e\Gamma (1-\b v_e\b V)$ where $\b v_e$ is the electron speed.
Furthermore,
we propose that the
Lorentz factor of electrons significantly exceeds the Lorentz factor of the
bulk motion, 
$\gamma \gg \Gamma$, and take into account that 
the up-scattered photon moves practically in the direction of the
electron (the accuracy of this approximation is of the order of
$1/\gamma$). Therefore, $\b v_e\approx \b n$ and, consequently,
$E'_e\approx E_e^{}/\dd$.

Integration of $w$ over all directions gives 
\[
W(E_e,\e_{\rm ph},\e)=\e^2\!\int\!w(\b p,\e_{\rm ph},\b k)\,d\Omega_e
\]
\be\label{com14}
 = \displaystyle\frac{8\pi r_e^2 }{E_e\,\eta}\left[ 2q\,\ln q+(1-q)\left(1+2q+
\frac{\eta^2q^2}{2\,(1+\eta q)}\right)\right],
\ee
where
\be
\eta=\frac{4\,\e_{\rm ph} E_e}{m^2}\,,\qquad
q=\frac{\e}{\eta\,(E_e-\e)}\,.
\ee
For the given values of 
$\e_{\rm ph}$ and $E_e$, the maximum energy of the upscattered photon 
is 
\be
\e_{\max}^{}=\frac{E_e}{1+1/\eta}\,.
\ee

Usually Eq.~(\ref{com14}) is obtained by integration over the directions of 
the momentum of the upscattered photon (\cite{Jones68} and 
\cite{Aharonian81}).
However, the argument that depends on the angle enters in 
the integrand as $(\b k\b p)$, therefore 
there is no difference over the directions of which vectors, $\b k$ or $\b p$,
the integration is performed. The above formulae 
are applicable under the conditions 
 \be
 \e_{\rm ph} \ll m \ll E_e\,.
 \ee

Let's denote by $\widetilde W(E_e,\e)=\frac{dN_{\gamma}}{dt\, d\e}$ the 
upscattered photon energy spectrum per electron: 
\be\label{com15}
\widetilde W(E_{e},\e)=\int\! W(E_e,\e_{\rm ph},\e)\,
n_{\rm ph}(\epsilon_{\rm ph}) \,d\e_{\rm ph}\,.
\ee
For the above simplifications, the observed energy flux ($F_e=\e^3\,\t g$)
becomes
\be\label{ECflux}
F_\e=\frac{\e\dd^3}{r^2}\int\!
N'_{e}\!\left(\frac{E_{e}}{\dd}\right)\,\widetilde W(E_e, \e)\, dE_{e}\,,
\ee
where $N'_{e}(E'_e)=E^{\prime 2}_e\int\! f'(E'_e,\b r'_0)\,d^3r'_0 $ is 
the differential number of electrons in the comoving
frame per energy and per solid angle. In the Thomson limit ($\eta\ll 1$),
$\e\,\widetilde W$ becomes a function of a single argument, $\e/E_e^2$. Writing
$\e\,\widetilde W(E_e,\e)=\Phi(\e/E_e^2)$, and performing an integration over
$E'_e=E_e/\dd$, we obtain
\be\label{ECflux1}
F_\e=\frac{\dd^4}{r^2}\int\!
N'_{e}\!\left(E'_e\right)\Phi\!\left(\frac{\e}{\dd^2E_e^{'2}}\right) dE'_e\,.
\ee
This implies that if the EC flux of a source at rest is described by 
some function $S(\e)$, the relativistic bulk motion of the source results in 
\be\label{ECflux2}
F_\e=\dd^4\,S(\e/\dd^2)\,.
\ee
In a log-log plot the function $F_e$ is obtained from $S$ by moving the latter 
up by a factor of $\log_{10}(\dd^4)$, and shifting it to the right by 
a factor of $\log_{10}(\dd^2)$. The total intensity is enhanced by a  factor of
 $\dd^6$:
\be\label{ECflux3}
\int_0^\infty \!F_\e\,d\e=\dd^6\int_0^\infty\!S(\e)\,d\e\,.
\ee
Integrating over all angles, we obtain the luminosity detected by the
observer (the apparent luminosity):
\be\label{ECflux4}
L=\frac15\,(16\,\Gamma^4-12\,\Gamma^2+1)\,L_0\,,
\ee
where $L_0$ is the luminosity of the source at rest (the intrinsic luminosity).

The above results are obtained for an arbitrary distribution of relativistic
elections. For electrons with a power-law distribution, 
$N_e'(E_e') \propto E_e'^{-p}$, in the Thomson limit 
we arrive at the well known result for the beaming pattern,
$F_{\epsilon_{\gamma}} \propto \dd^{3+p}\,$ \citep{Dermer95}. 

\section{Non-isotropy}\label{non_isotropy}

In this section we study the impact of a possible anisotropy of the electron
distribution inside the moving source on the angular and energy distributions of
EC radiation. Non-isotropic distribution of electrons might be caused by
different reasons. In particular, anisotropic particle distributions are
expected within different acceleration scenarios, including the particle
acceleration by relativistic shocks (see e.g. \citealt{Dempsey07}), by the
converter mechanism \citep{Derishev03} or due to magnetic reconnection
\citep{Cerutti12}. 

We will use the general Eq.~(\ref{com12}) assuming that the source function
(distribution of electrons) in the comoving system is not isotropic, but
stationary. Let's introduce anisotropy (in $K'$ ) in the form 
\be\label{anis0}
f'(\b p',\b r')=\Psi(\b n'_e)\,f'(E'_e,\b r') \,,
\ee
where $\b n'_e=\b p'/|\b p'|$. The calculations for anisotropy in a general
form are quite complex, therefore, for the  purpose of demonstration, we will use an
empirical approach, namely, adopt  the specific form of elongated ellipsoid of
revolution described by the function
\be\label{anis1}
\Psi(\b n'_e)=
\frac{\sin\alpha}{\alpha\sqrt{1-(\b s'\b n'_e)^2\sin^2\alpha}}\,.
\ee\
Here $\b s'$ denotes a constant unit vector, which could have an arbitrary
direction, and the parameter $\alpha$ is confined within the interval
$0<\alpha<\pi/2$. The function $\Psi$ is normalized by the condition 
\be\label{anis2}
\int\!\Psi(\b n'_e)\,\frac{d\Omega'_e}{4\pi}=1\,.
\ee
The angular distribution given by Eq.~(\ref{anis1}) is azimuthally symmetric 
relative to $\b s'$. If the axis $z$ is directed along $\b s'$, the function 
$\Psi(\b n'_e)$ can be considered as an equation of the surface of the ellipsoid
of revolution written in spherical coordinates. The semi-major axis of the
ellipsoid is directed along $\b s'$ and is equal to
$a_>=\frac{\tan\alpha}{4\pi\alpha}$; the another two semi-axes 
are $a_<=\frac{\sin\alpha}{4\pi\alpha}$.

It is convenient to introduce the asymmetry parameter $\lambda$, 
\be\label{anis3}
\lambda=\frac{a_>}{a_<}-1=\frac1{\cos\alpha}-1\,.
\ee
In the limit of $\lambda\to 0$, the ellipsoid degenerates into a sphere, i.e.
the angular distribution becomes isotropic ($\Psi\big|_{\lambda\to 0}=1$). At 
$\lambda\gg 1$, the angular distribution is strongly extended in the directions
of $\b s'$ and $-\b s'$. We allow $\lambda$ to be function of $E'_e$, i.e. 
the anisotropy can be energy-dependent.

In the case of anisotropic angular distribution, the integration of
Eq.~(\ref{com14}) over directions of the electron momentum results in 
\be\label{anis4}
W_{\rm anis}=\e^2\!\int\!\Psi(\b n'_e) w(\b p,\e_{\rm ph},\b k)\,d\Omega_e.
\ee
Since the function $w$ is different from zero when the directions 
of vectors $\b p$ and $\b k$ practically coincide, the argument of the
function $\Psi$ in Eq.~(\ref{anis4}) can be replaced by $\b n'$, and $\Psi(\b
n')$ can be taken out under the sign of integral. This gives 
\be\label{anis5}
W_{\rm anis} \approx \Psi(\b n')\, W(E_e,\e_{\rm ph},\e)\,,
\ee
where $\b n'$ in the argument of $\Psi$ should be expressed via $\b n$ 
according to relations in Eq.~(\ref{com11}). Therefore, for the anisotropic 
angular distribution of electrons the observed flux is
\be\label{ECflux2}
F_\e=\frac{\e\dd^3}{r^2}\int\!\Psi(\b n')\,
N'_{e}\!\left(\frac{E_{e}}{\dd}\right)\,\widetilde W(E_e, \e)\, dE_{e}\,,
\ee
It is interesting to compare this equation with Eq.~(\ref{ECflux}).
In the Thomson regime of scattering, Eq.~(\ref{ECflux2}) is simplified, 
\be\label{ECflux3}
F_\e=\frac{\dd^4}{r^2}\int\!\Psi(\b n')\,
N'_{e}\!\left(E'_e\right)\Phi\!\left(\frac{\e}{\dd^2E_e^{'2}}\right) dE'_e\,.
\ee
In the case of an energy-independent anisotropy ($\lambda= \rm const$), for 
the total (integrated over energy) intensity of radiation in the given
direction, we have 
\be\label{D6}
 \int\! F_\e\,d\e\propto \Psi(\b n')\,\dd^6 \ .
\ee

For illustration of the results, we fix the orientations of the axes in the
following way. As before, the axis $z$  is directed  along the velocity $V$; the
axis $y$ we choose from the condition that the vector $\b s'$ becomes parallel to
the $(y,z)$ plane. Then the components of the vector $\b s'$ can be written in
the following form 
\be
\b s'=(0,\sin\theta'_*,\cos\theta'_*)\,.
\ee
Here the angle $\theta'_*$ determines the direction of the axis of symmetry of the 
angular distribution of electrons in the comoving frame.

To simplify the analysis, we assume that the radiation is detected in the plane
$(y,z)$, and introduce polar coordinates in this plane. Then $(\b s'\b
n')=\cos(\theta'-\theta'_*)$, where $\theta'$ varies in the interval from $-\pi$
to $\pi$ ($\theta'=0$ for the points on the axis $z$). Expressing in this
equation $\theta'$ through the viewing angle $\theta$ in the system $K$, we get
\be
(\b s'\b n')= \dd\left(\sin\theta\sin\theta'_*+\Gamma(\cos\theta-V)\cos\theta'_*
\right).
 \ee
This expression is to be substituted into Eq.~(\ref{anis1}).

\begin{figure*}
 \centering
 \includegraphics[width=0.31\textwidth,angle=-90]{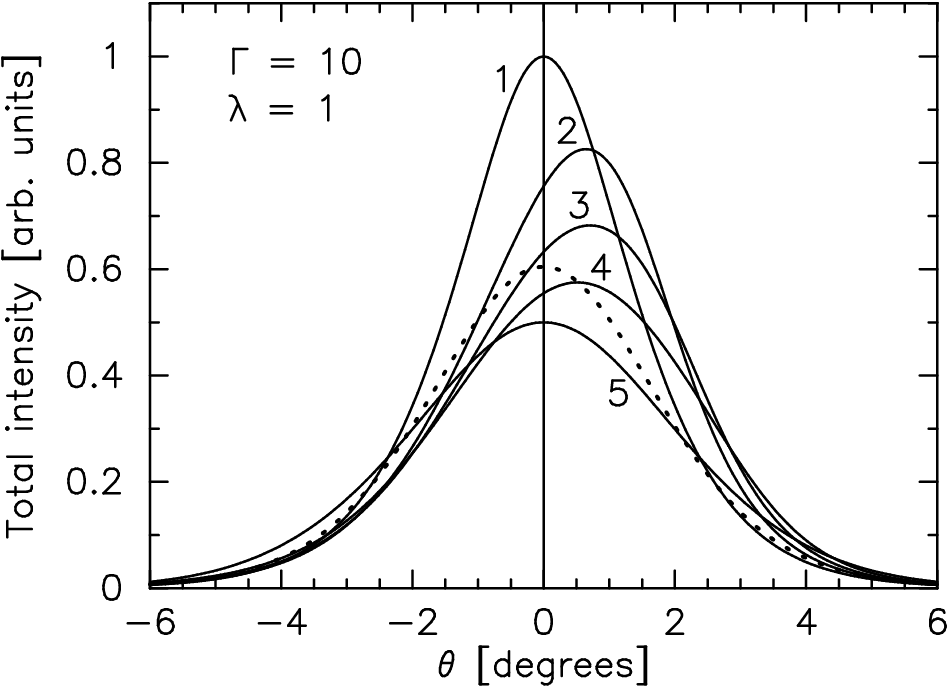}\quad
 \includegraphics[width=0.31\textwidth,angle=-90]{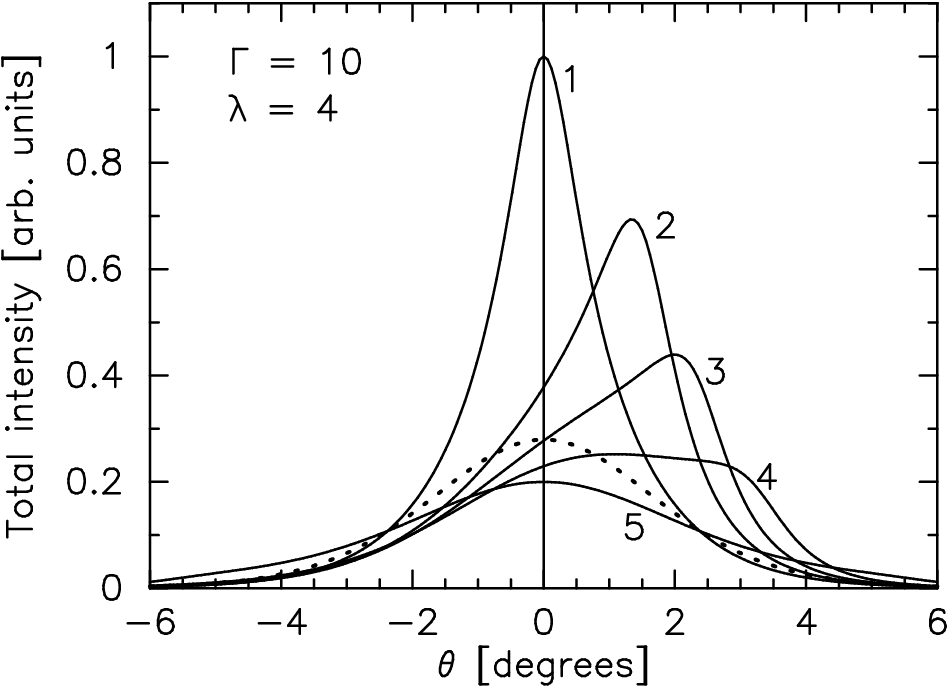}\\[8pt]
 \includegraphics[width=0.31\textwidth,angle=-90]{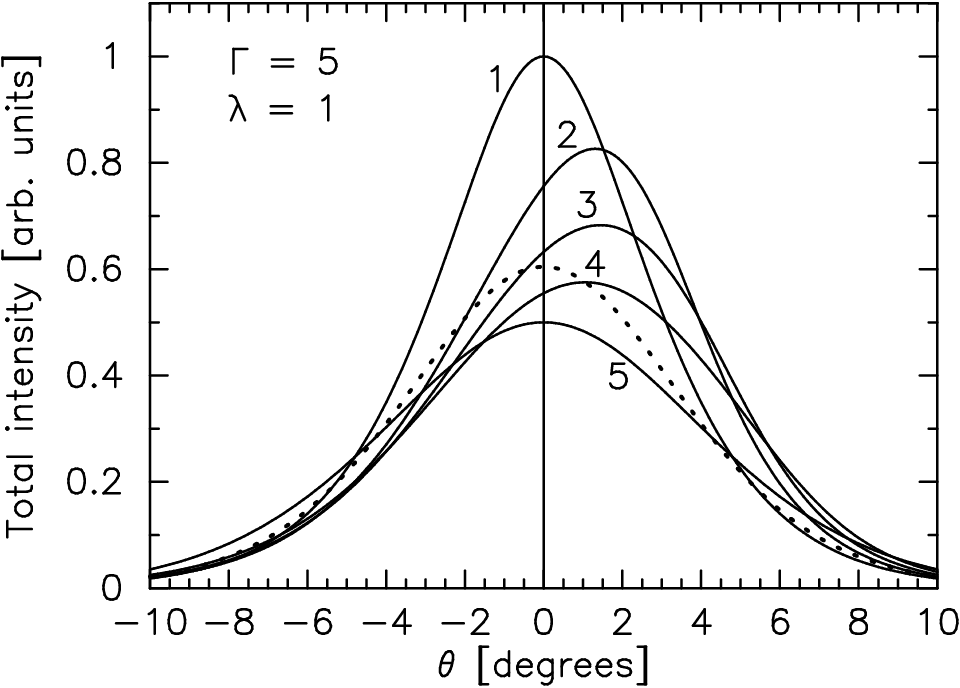}\quad
 \includegraphics[width=0.31\textwidth,angle=-90]{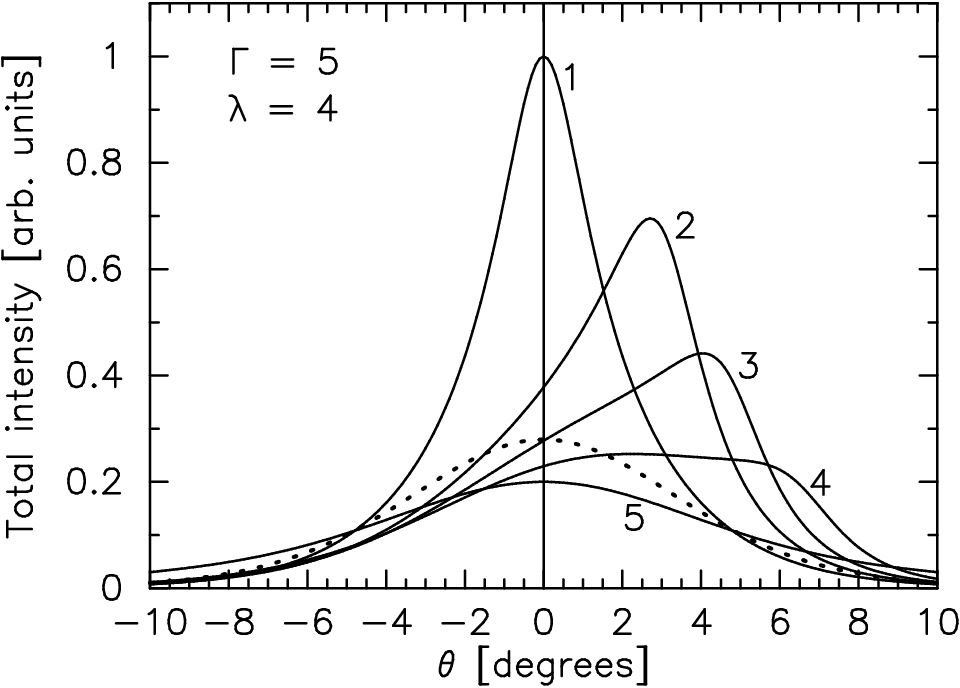}
 \caption{The intensity of the EC radiation detected by an observer as a
function of  the angle between the line of sight and the direction of the
relativistically moving source $\theta$. Calculations are performed for four
different combinations of the jet's Lorentz factor $\Gamma$ and the asymmetry
parameter $\lambda$ characterizing the angular distribution of electrons. The
curves correspond to different directions of the axis of symmetry of the angular
distribution of electrons in the comoving system $\theta'_*$: $0^\circ$ (curve
1), $30^\circ$ (curve 2), $45^\circ$ (curve 3), $60^\circ$ (curve 4), and
$90^\circ$ (curve 5). For comparison the intensity corresponding to the
isotropically distributed electrons is also shown (dotted curves).} 
\label{angle_dist}
 \end{figure*}

In Fig~\ref{angle_dist} we show the dependence of the intensity 
of the total EC radiation (the flux integrated over the photon energies) on 
$\theta$ - the angle between the line of sight and the direction of the jet. 
The calculations are performed for different combinations of the bulk motion
Lorentz factor $\Gamma$ and the asymmetry parameter $\lambda$. The curves shown
in these figures correspond to five  different directions of the axis of symmetry of the
electrons angular distribution described by the angle $\theta'_*$. For
comparison we show also the intensity for isotropically distributed electrons. 
One can see that except for $\theta'_*=0$ and $\theta'_*=\pi/2$, the maximum of
the observed radiation appears not at $\theta=0$, as it happens in the isotropic
case. Instead, it is shifted to larger angles as the level of anisotropy
increases. For the chosen parameter, the shift can be as large as several
degree. This implies that in the case of anisotropic distributions of electrons
in the source, we should be able to see misaligned jets as it has been indicated
by \cite{Derishev07}.

In a more general approach, the $\lambda$-parameter can depend on the electron
energy. This could be realized, for example, in the case of diffusive shock
acceleration of electrons when the low-energy particles can be effectively
isotropised in the downstream region due to pitch-angle scattering, whereas
higher-energy particles radiate away their energy before being fully isotropised
(see, e.g. \citealt{Derishev07}). To demonstrate this effect, let's assume a
simple energy-dependence of the asymmetry parameter, $\lambda=\lambda'
\gamma'^{p_1}$, and consider  a  distribution of electrons in the standard 
``power-law with an exponential cutoff'' form: 
\be\label{anis6}
N'_e(\gamma',\b n') = A\, \Psi(\b n')\,\gamma'^{-p}\,e^{-\gamma'/\gamma'_0} \ ,
\ee
where $A={\rm const}$. Here instead of the electron energy $E'_e$, 
we use its Lorentz factor, $\gamma'=E'_e/m_e$. From the condition of
normalization in Eq.~(\ref{anis2}) it  follows that $\int\!N'_e(\gamma',\b n')\,
d\Omega_{\b n'}$ does not depend on the level of anisotropy. 

\begin{figure}
 \centering
 \includegraphics[width=0.33\textwidth,angle=-90]{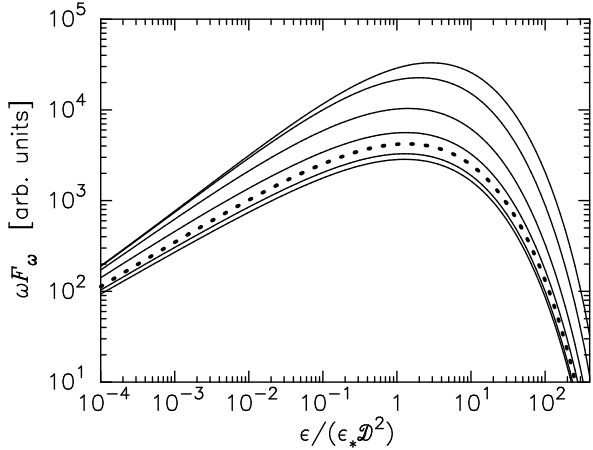}
 \caption{ The spectral energy distribution of the EC radiation 
 for different orientations of the axis of symmetry of angular distribution of
electrons.  The solid curves correspond to the angles 
$\theta'_*=0^\circ$, $5^\circ$, $15^\circ$, $30^\circ$, $60^\circ$, $90^\circ$
(from top to bottom). The dotted line is the spectrum of radiation corresponding
to the isotropic distribution of electrons. The energy is expressed in the
units $\e_*=4\,kT\,{\gamma'_0}^2$.}
\label{EC_anis}
 \end{figure}

In Fig.~\ref{EC_anis} we show the spectral energy distribution of 
EC radiation assuming that a relativistic jet propagates through the 2.7~K
microwave background radiation. This might be relevant to the inverse Compton
X-ray emission of extended jets of AGN  which could be (still) relativistic on
kpc scales (see, e.g. \citealt{Sambruna2002}). The radiation spectra are
calculated for the viewing angle $\theta=0$, thus $\dd \approx 2 \Gamma$, and
for the following combination of model parameters: $p=2$, $\lambda'=0.1$,
$p_1=1/2$ and $\gamma'_0=10^4$. Since $\eta\sim 4\,kT\,\gamma'_0/m\approx
2\times 10^{-5}\ll 1$, the Compton scattering proceeds in the Thomson limit. One
can see from 
 Fig.~\ref{EC_anis} that if the photon energy is expressed in units of $\dd^2
\e_*=4\,kT\,{\gamma'_0}^2 \dd^2$, the shape of the energy spectrum of EC
radiation does not depend on the jet's Doppler factor. On the other hand, it
depends on the angle $\theta'_*$. The apparent reason is the dependence of the
electron energy distribution on $\theta'_*$. For example, for $\theta=0$ and
$\lambda'\,{\gamma'}^{p_1}\gg 1$, the spectrum given by Eq.~(\ref{anis6}) for 
$\theta'_*=0$ can be written in the form 
\be\label{anis7}
N'_e(\gamma',\b n') =
\frac{\lambda' A}{2\,\pi^2}\,\gamma'^{-p+p_1}\,e^{-\gamma'/\gamma'_0}\,,
\ee
For $\theta'_*=\pi/2$, Eq.~(\ref{anis6}) differs from the spectrum
corresponding to the isotropic distribution of electrons, by a constant factor
of $2/\pi$.

\section{Summary} 

In various astrophysical objects, like microquasars, AGN, and GRBs, the observed
fluxes of radiation emerge from relativistically moving jets. The Doppler
boosting caused by this motion can significantly enhance (by orders of
magnitude!) the emitted absolute flux and shift the spectrum towards higher
energies. Therefore, the beaming pattern of radiation is a key issue for proper
understanding of acceleration and radiation processes in these objects. 

In this paper we derived the energy distribution of the EC radiation by solving
the photon transfer equation for an optically thin source in a rather general
case. It is described by Eq.~(\ref{com12}) which allows non-stationary and
non-isotropic distribution of electrons in the frame of a relativistically
moving source. Eq.~(\ref{com12}) does not specify the energy distribution of
electrons either, but requires isotropic, homogeneous and non-variable fields of
seed photons for ICS. The latter condition makes the solution independent of the
spatial distribution of electrons. For power-law energy distribution of
isotropically distributed electrons Eq.~(\ref{com12}) is reduced to previously
derived results \citep{Dermer95,Kirk01}. 

Anisotropic distribution of electrons in a relativistically moving source can be
realized in some  acceleration scenarios, therefore it is of a special practical
interest. The formalism developed in this paper has been used to study the
impact of the electron anisotropy on the EC emission. The calculations show that
the anisotropy of emitting particles can significantly modify the beaming
pattern. Most notably, the emission peak can be significantly shifted relative
to the line of sight. This implies that, thanks to the anisotropic distribution
of electrons in the source, modestly mis-aligned jets may become detectable.
And vice versa, while the energy of strongly anisotropic distribution of
electrons in a source at rest can be radiated away from the observer, the
relativistic motion of the source would make the radiation detectable, even in
the case of most unfavorable anisotropy of electrons. 

Generally, the electron anisotropy is expected to be energy-dependent. In this
case the anisotropy could result in harder spectra of EC emission compared to
the isotropic distribution of electrons. 
The effects related to the anisotropic distribution of electrons in general, and
in the context of the EC scenario, in particular, are quite strong. They cannot
be ignored when interpreting the high energy emission from highly relativistic
jets in AGN and GRBs. 


\end{document}